\documentstyle[12pt]{article}

\def\C{{\rm\kern.24em \vrule width.02em height1.4ex depth-.05ex \kern-.26em
C}}
\def\P{{\rm I\kern-.20em P}}
\def\pa{\partial}

\newcommand{\beq}{\begin{equation}}
\newcommand{\eeq}{\end{equation}}

\begin{document}
\baselineskip=20pt
\pagestyle{plain}

\title{Equivalence of Geometric $h<1/2$ and Standard \\
$c>25$ Approaches \\
to Two-Dimensional Quantum Gravity}
\author{
Leon A.~Takhtajan \\
Department of Mathematics\\
SUNY at Stony Brook\\
Stony Brook, NY 11794-3651 \\
U.S.A.}
\date{September 6, 1995}
\maketitle

\begin{center}
{\bf Abstract\footnote{hep-th/9509026}}
\end{center}
\begin{quote}
We show equivalence between the standard weak coupling regime $c>25$ of
the the two-dimensional quantum gravity and regime $h<1/2$ of the original
geometric approach of Polyakov~\cite{Pol1,Pol2}, developed in~\cite{TC,T,TV}.
\end{quote}

 {\bf 1} In this letter I shall demonstrate the equivalence of two approaches
to the two-dimensional quantum gravity. The first approach, called geometric,
goes back to Polyakov's original discovery~\cite{Pol1}, and was presented
in~\cite{Pol2}. According to~\cite{Pol2}, correlation functions of the
Liouville vertex operators are represented by the functional integral with the
Liouville action over all Riemannian metrics in a given conformal class with
prescribed singularities at insertion points. This approach can be viewed
as quantization of the hyperbolic geometry in two dimensions, and for the
case of puncture operators, it has been extensively developed in our
papers~\cite{TC,T,TV}. In these papers, we have computed conformal weights of
puncture operators, the central charge, and also analyzed semi-classical
behavior of the theory and its non-trivial relation with the Weil-Petersson
geometry of Teichm\"{u}ller spaces. Notably, the latter provides the
Friedan-Shenker ``program'' of modular geometry~\cite{FS} with a meaningful
example. It is also worthwhile to mention that the first validation of
geometric approach was provided by a rigorous proof~\cite{ZT} of the
Polyakov's conjecture~\cite{Pol2} that the classical Liouville action solves
the problem of accessory parameters for the Fuchsian uniformization of Riemann
surfaces, posed by Klein~\cite{K} and Poincar\'{e}~\cite{P1}.

Another approach to the Liouville theory, which we call standard, goes back
to Braaten-Curtright-Thorn~\cite{CT} who have computed conformal weights and
the central charge of the theory using canonical quantization. Subsequently,
these results were obtained by Knizhnik-Polyakov-Zamolodchikov~\cite{KPZ},
using methods of conformal field theory in the light-cone gauge, and by
David-Distler-Kawai~\cite{DDK} in the conformal gauge. The connection between
this approach and the hyperbolic geometry, uniformization and spectral theory
of automorphic functions is less transparent than for the geometric one. On
the other hand, standard approach allows one to use powerful methods of
conformal field theory, such as free fields representation and analytic
continuation, to study correlation functions of the theory.

Till now, these two approaches have looked look quite different and relation
between them seemed obscure. In particular, geometric approach gives the
following value to the central charge of the theory
\beq \label{ccg}
c_{geom}=1+\frac{12}{h},
\eeq
with $h>0$ being a bare coupling constant, whereas in standard approach
\beq \label{ccs}
c_{stand}=1+6(b+1/b)^2,
\eeq
where $b$ is a corresponding Liouville coupling. Thus $c_{stand}>25$, while
$c_{geom}$ takes all values greater than $1$. Moreover, in geometric
approach conformal weights of geometric vertex operators (i.e.~that with
charges corresponding to the Fuchsian uniformization) remain classical.

Recently, there was published a remarkable paper by A.~and Al.~Zamolodchikov's~
\cite{ZZ}, where they provided an elegant geometric setting for standard
approach and presented convincing arguments for Dorn-Otto conjecture~\cite{DO}
on structure constants for the operator algebra of the Liouville theory.
Using formulation in~\cite{ZZ}, it is possible to show that methods developed
in~\cite{T,TV} work for standard approach as well. In particular, as we
shall demonstrate here, standard values for the conformal weights can be
obtained by pure geometric arguments, based on the exact form of the Liouville
action. Formula (\ref{ccs}) for the central charge can be also derived in the
same manner as (\ref{ccg}).

Moreover, these two apparently different approaches turn out to be equivalent
in weak coupling regime $h<1/2$ and $c>25$! In order to understand this, we
recall that dependence of physical parameters of the theory like conformal
weights, central charge, etc., on the bare couplings is irrelevant and only the
relation between them is fundamental. Therefore, one has a freedom of changing
couplings, rescaling fields, etc., which in our case can be used as follows.
First, start with the geometric approach and investigate whether it admits
vertex operators with conformal weight $1$. It turns out that for weak
coupling regime $h<1/2$ one has two such operators $V_{1,2}$ with
charges
$$\alpha_{1,2}(h)=\frac{1 \mp \sqrt{1-2h}}{h}.$$
Second, introduce a new bare coupling constant
$$b(h)=\sqrt{\frac{h}{2}}\alpha_1(h)=\frac{1-\sqrt{1-2h}}{\sqrt{2h}},$$
so that $b(h) \rightarrow 0$ as $h \rightarrow 0$. As we shall show in the
main text, formulas for conformal weights and central charges for geometric
and standard approaches with corresponding coupling constants $h$ and $b(h)$,
become identical in weak coupling regime $h<1/2$.

The paper is organized as follows. In the second section we briefly recall
basic facts of geometric approach and present the geometric derivation
of conformal weights. In the third section we summarize, in similar
fashion, standard approach, and in the last section we elaborate the
arguments for their equivalence in the weak coupling regime.

{\bf 2} Here we outline main principles of geometric approach (see
\cite{TC,T,TV} for details).  Let $X$ be Riemann sphere $\P^{1}$ with
$n$ marked distinct points $z_1, \ldots, z_{n-1},z_n=\infty$. According
to~\cite{Pol2,TC,T,TV}, the $n$-point correlation function of Liouville vertex
operators $V_{\alpha}(z)=e^{\alpha \phi(z,\bar{z})}$---spinless primary fields
of the theory---is defined by the following functional integral
\beq \label{correlation}
<V_{\alpha_1}(z_1) \cdots V_{\alpha_{n-1}}(z_{n-1})V_{\alpha_n}(\infty)>=
{\bf \int}_{{\cal C}(X)}{\cal D}\phi~e^{-(1/2\pi h)S_{X}(\phi)},
\eeq
where $h$ plays the role of Liouville coupling constant. Here ``domain of
integration'' ${\cal C}(X)$ consists of all smooth conformal metrics
$ds^2=e^{\phi(z, \bar{z})}|dz|^{2}$ on $X$ satisfying the following asymptotics
\beq \label{asymptotics}
\phi(z,\bar{z})\simeq - \alpha_i h \log|z-z_i|^2,~z \rightarrow z_i,~i=1,
\ldots, n-1,
\eeq
and
\beq \label{infinity}
\phi(z,\bar{z})\simeq (\alpha_n h-2) \log|z|^2,~z \rightarrow z_n=\infty.
\eeq
where $z=x+\sqrt{-1}y$ is a complex coordinate on $\C=\P^1 \setminus
\{\infty \}$.  Charges $\alpha_i$ satisfy restriction $\alpha_i h <1$
(cf.~\cite{S}), which is equivalent to the finiteness of the area of $X$ with
respect to metric $ds^2$. In the limiting case $\alpha_i=1/h$, which
corresponds to the punctures, asymptotics
(\ref{asymptotics})--(\ref{infinity}) should be modified
\beq \label{punctures}
\phi(z,\bar{z}) \simeq -\log|z-z_i|^2 - \log\log^{2}|z-z_i|,~z \rightarrow
z_i,~i=1, \ldots, n-1,
\eeq
and
\beq \label{pinfinity}
\phi(z,\bar{z}) \simeq -\log|z|^2 - \log\log^{2}|z|,~z \rightarrow z_n=\infty,
\eeq
so that the total area of $X$ remains finite. The action functional is given by
$$S_{X}(\phi)=\lim_{\epsilon \rightarrow 0}S_{\epsilon}(\phi),$$
where
\begin{eqnarray}
S_{\epsilon}(\phi) &=& \int \int_{X_{\epsilon}}(|\phi_{z}|^{2}
+e^{\phi})dx \wedge dy - \frac{h\sqrt{-1}}{2} \sum_{i=1}^{n-1} \alpha_i
\int_{\gamma_i} \phi(\frac{d \bar{z}}{\bar{z}-\bar{z}_i}-\frac{dz}{z-z_i}) \\
&-& \frac{\sqrt{-1}}{2}(\alpha_n h-2) \int_{\gamma_n}\phi(\frac{d\bar{z}}{\bar
{z}}-\frac{dz}{z})- 2 \pi h^2\sum_{i=1}^{n-1}\alpha^{2}_{i} \log \epsilon -
2 \pi (\alpha_n h -2)^2 \log \epsilon, \nonumber
\end{eqnarray}
where $X_{\epsilon}= X \setminus (\bigcup^{n-1}_{i=1}\{|z-z_i|<\epsilon\}
\bigcup \{|z|>1/\epsilon\})$ and
$$\gamma_i(t)=z_i+\epsilon e^{2 \pi\sqrt{-1}t},~i=1,
\ldots, n-1,~\gamma_n(t) =\frac{1}{\epsilon}e^{2 \pi \sqrt{-1}t},~0 \leq t
\leq 1.$$
 Note that in this form of the regularized action (cf.~\cite{ZT,T,ZZ}) line
integrals are necessary in order to ensure the proper asymptotic behavior
(\ref{asymptotics})--(\ref{infinity}). For the case $\alpha_i h=1$ one should
modify the line integrals by adding terms $dz/((z-z_i)\log|z-z_i|)$ and their
complex conjugates, as well as adding overall term $2\pi n \log| \log\epsilon|$
(cf.~\cite{ZT,T}).

Conformal weights $\Delta_{\alpha}$ of Liouville vertex operators
$V_{\alpha}(z)$ are given by the following formula
\beq \label{weight}
\Delta_{\alpha}=\frac{h}{2}\alpha(\frac{2}{h}-\alpha),
\eeq
which can be simply derived from the Liouville action~\cite{T,TV}. Namely,
consider three-point correlation function $<V_{\alpha_1}(z_1)V_{\alpha_2}
(z_2)V_{\alpha_3}(\infty)>$. Since, according to BPZ \cite{BPZ},
$$V_{\alpha}(\infty)=\lim_{z \rightarrow \infty}|z|^{4\Delta_{\alpha}}
V_{\alpha}(z),$$
this correlation function has the form
\beq \label{3-point}
<V_{\alpha_1}(z_1)V_{\alpha_2}(z_2)V_{\alpha_3}(\infty)>=\frac{C(\alpha_1,
\alpha_2, \alpha_3)}{|z_1-z_2|^{2\Delta_1+2\Delta_2-2\Delta_3}},
\eeq
where $\Delta_i=\Delta_{\alpha_i}$ and $C(\alpha_1,\alpha_2,\alpha_3)$ is
the structure constant of the operator algebra of the theory. On the other
hand, $<V_{\alpha_1}(z_1)V_{\alpha_2}(z_2)V_{\alpha_3}(\infty)>$ is
represented by a functional integral (\ref{correlation}) with $n=3$. From
the latter form it is easy to determine the coordinate dependence of the
correlation function using global conformal invariance. Namely, consider
fractional-linear transformation
$$\sigma z =\frac{z-z_1}{z_2-z_1},$$
which maps Riemann surface $X$ with three marked points $z_1,z_2,\infty$
onto normalized Riemann surface $\tilde{X}$ with three marked points $0,1,
\infty$. Since in geometric approach, $e^{\phi}$ transforms like $(1,1)$-tensor
under local change of coordinates, one has
\beq \label{1,1}
\tilde{\phi}(\sigma z)=\log |z_1-z_2|^2+\phi(z).
\eeq
Now, straightforward computation yields
\beq \label{transformation}
S_X(\phi)-S_{\tilde{X}}(\tilde{\phi})=2 \pi h(\Delta_1+\Delta_2-\Delta_3)
\log|z_1-z_2|^2,
\eeq
where $\Delta_{\alpha}=-h\alpha^2/2 + \alpha$. Note that the first term in this
formula for $\Delta_{\alpha}$, which represents a ``free-field contribution'',
comes from the transformation property of the surface and line integrals in the
action (8). The second term, which equals to the classical conformal weight of
$V_{\alpha}(z)$, comes entirely from the line integrals and the transformation
law (\ref{1,1}). Finally, observing that the local change of variables
$\phi(z) \mapsto \phi(\sigma z)$ in the functional integral (\ref{correlation})
leaves ``integration measure'' ${\cal D}\phi$ invariant, we get
(\ref{3-point}), where conformal weights are given by (\ref{weight}) and
$$C(\alpha_1,\alpha_2,\alpha_3)=<V_{\alpha_1}(0)V_{\alpha_2}(1)V_{\alpha_3}
(\infty)>.$$

Note that conformal weights are invariant with respect to reflection
$\alpha \mapsto 2/h-\alpha$ with the fixed point being $\alpha=1/h$. For such
$\alpha$ one can define puncture operator as
$$P(z)=\phi(z)V_{1/h}(z)=\frac{\pa}{\pa \alpha}V_{\alpha}(z)|_{\alpha=1/h},$$
(cf.~\cite{ZZ,S}) according with the double logarithm term in asymptotics
(\ref{punctures})--(\ref{pinfinity}).

It should be also noted that there is a discrete series of charges
$$\alpha_l=\frac{1}{h}(1-\frac{1}{l}),~l~{\rm an~integer}>1~{\rm or}~l=\infty,
$$
which correspond to the Fuchsian uniformization of Riemann surface $X$
with elliptic fixed points of finite order ($l < \infty$), or with punctures
($l=\infty$). For such $\alpha$'s, which correspond to geometric vertex
operators, formula (\ref{weight}) gives the following conformal weights
$$\Delta_l=\frac{1}{2h}(1-\frac{1}{l^2}).$$
It is remarkable that these weights coincide (times $1/h$) with the
uniformization data, given by the coefficients at the second order poles of the
Schwarzian derivative of the inverse function of the uniformization map of
$X$~\cite{K,P1}. The latter quantity, according to~\cite{P2}, coincides with
classical stress-energy tensor $T_{cl}= T(\phi_{cl})$, where
\beq \label{set}
T(\phi)=\frac{1}{h}(\phi_{zz}-\frac{1}{2}\phi_{z}^{2}).
\eeq
(See \cite{TC,TV} for more details and references).

Finally, the central charge of the Liouville theory can be computed from the
short-distance behavior of the two-point correlation function of the
stress-energy tensor in the presence of vertex operators
\beq \label{TT}
<T(z)T(w)V_{\alpha_1}(z_1) \cdots V_{\alpha_n}(\infty)>={\bf \int}_{{\cal
C}(X)}
{\cal D}\phi~T(\phi)(z)T(\phi)(w)~e^{-(1/2 \pi h)S_X(\phi)}.
\eeq
Calculation from \cite{T,TV} (performed for the case of puncture operators)
gives the following expression for the central charge
\beq \label{central charge}
c_{geom}=1+\frac{12}{h},
\eeq
where the only quantum correction to the semi-classical value $12/h$
comes from the one-loop contribution and is equal to $1$.

We will not dwell here upon the relation of geometric approach to the
Friedan-Shenker modular geometry and Weil-Petersson geometry of Teichm\"{u}ller
spaces, referring to \cite{TC,T,TV}.

{\bf 3} Here we briefly recall, following \cite{ZZ}, the basic facts from
standard approach. The correlation function of vertex operators
$e^{2a\phi}$ can be defined by the following functional integral
\beq \label{tcorrelation}
<e^{2a_1 \phi(z_1)} \cdots e^{2a_n \phi(z_n)}>_{Q}={\bf \int}_{
{\cal C}(X,Q)}{\cal D}\phi~e^{-A_{X,Q}(\phi)},
\eeq
where ``domain of integration'' ${\cal C}(X,Q)$ now consists of all
$(Q/2,Q/2)$-tensors $e^{\phi(z, \bar{z})}$ on $X$ having the asymptotics
(\ref{asymptotics}) (with the replacement $n-1 \mapsto n$ and $\alpha \mapsto
a$) and having the ``charge $Q$'' at infinity
\beq \label{tinfinity}
\phi(z,\bar{z}) \simeq -Q \log|z|^2,~z \rightarrow \infty.
\eeq
The action is given by
$$A_{X,Q}(\phi)=\lim_{\epsilon \rightarrow 0}A_{\epsilon}(\phi),$$
where
\begin{eqnarray}
A_{\epsilon}(\phi) &=& \frac{1}{\pi} \int \int_{X_{\epsilon}}
(|\phi_{z}|^{2} +\pi\mu e^{2b\phi})dx \wedge dy + \frac{\sqrt{-1}}{2 \pi}
\sum_{i=1}^n a_i \int_{\gamma_i} \phi(\frac{d \bar{z}}{\bar{z}-\bar{z}_i}-
\frac{dz} {z-z_i}) \\ \nonumber
&+& \frac{\sqrt{-1}Q}{2 \pi}
\int_{\gamma_{\infty}}\phi(\frac{d\bar{z}}{\bar{z}}
-\frac{dz}{z})-2(\sum_{i=1}^n a_{i}^2 + Q^2) \log \epsilon,
\end{eqnarray}
with $\mu$ being a cosmological constant and $b$ being a Liouville coupling
constant.

Analyzing the three-point correlation function using the arguments from the
previous section and using the new transformation law for $e^{\phi}$, one gets
the following expression for conformal weights $\Delta_a$ of vertex operators
$e^{2a \phi(z)}$:
\beq \label{tweight}
\Delta_a=a(Q-a).
\eeq
The stress-energy tensor of the theory has the form
\beq \label{tset}
T_Q(\phi)=-\phi_{z}^{2}+Q \phi_{zz},
\eeq
and the same arguments as in~\cite{T,TV} give the following formula for the
central charge

\beq \label{Q}
c_{stand}=1+6Q^2.
\eeq

To relate global charge $Q$ and Liouville coupling constant $b$, one
imposes the condition that the ``perturbation'' operator $e^{2b\phi(z)}$ has
conformal weight $1$. It results in
\beq \label{Qb}
Q=b+1/b,
\eeq
and gives standard restriction $c>25$ on the central charge of the theory
in the weak coupling regime (which corresponds to real $b$). For further
discussion of this approach, and correlation functions properties in
particular, we refer to~\cite{ZZ,DO} and references therein.

Arguments presented here are nothing but geometric interpretation of KPZ-DDK
results. Note that accurate (cf.~\cite[Sect.~3.7]{GM}) definition of the
Liouville action requires line integrals terms, which play fundamental
role in geometric derivation of conformal weights. Being linear in the
Liouville field, these terms do not contribute to the perturbation theory.

{\bf 4}. Here we compare both approaches and establish their equivalence for
weak coupling regime $h<1/2$ and $c>25$.

At first glance, these approaches look strikingly different. Indeed, one
has $c_{stand}>25$, whereas $c_{geom}>1$. What is more, in standard
approach, operator $e^{b\phi(z)}$ has conformal weight $1$, whereas in
geometric approach conformal weight of $V_{1}(z)=e^{\phi(z)}$ is $1-h/2$. The
latter even looks like a drawback of the geometric approach. However, that is
not really so, since in standard approach, classical condition of $e^{\phi}$
being a $(Q/2,Q/2)$-tensor is also ``violated'' after quantization: $\Delta_{
1/2}=Q/2-1/4 \neq Q/2$. Nevertheless, there is nothing wrong with these
results, since in the formalism of functional integration, one should
integrate over all classical configurations, whether they are conformal
Riemannian metrics in geometric approach, or $(Q/2,Q/2)$-tensors in standard
approach.

Still, the importance of the latter remark is in the fundamental role played
by vertex operators of conformal weight $1$. It turns out that in the
weak coupling regime of geometric approach, it is also possible to find
such operators. Simply solving equation $\Delta_{\alpha}=1$, we get the
roots
$$\alpha_{1,2}=\frac{1 \mp \sqrt{1-2h}}{h},$$
which are real if and only if $h<1/2$. In this regime, $c_{geom}>25$, which
suggests the equivalence with standard approach.

Indeed, starting from geometric approach and setting
$$Q(h)=\sqrt{\frac{2}{h}},~\alpha=Qa$$
(thus effectively rescaling Liouville field $\phi$) we get from (\ref{weight})
and (\ref{central charge}) formulas (\ref{tweight}) and (\ref{Q}):
$$\Delta_a=\Delta_{\alpha}=a(Q-a),~~c_{stand}=1+6Q^2.$$
Moreover, real roots $\alpha_{1,2}$ satisfy $\alpha_1 \alpha_2 = Q^2,~\alpha_1+
\alpha_2=Q^2$, so that $a_1 a_2=1,~a_1+a_2=Q$. Thus, introducing
$$b(h)=a_1=\frac{1-\sqrt{1-2h}}{\sqrt{2h}},$$
which is real for $h<1/2$, we get the constraint (\ref{Qb}):
$$Q(h)=b(h)+1/b(h),$$
which establishes the equivalence between geometric and standard approaches.

It is well known that it is extremely difficult to extend standard approach
to strongly coupled regime $c<25$ (formally coupling constant $b$ in
(\ref{ccs}) becomes pure imaginary). On the other hand, geometric approach
seems to be well-defined for all positive values of $h$, and in particular,
for strongly coupled regime $h \geq 1/2$, for which $c_{geom} \leq 25$. It
suggests that this regime of geometric approach can be considered as strongly
coupled regime for standard approach. Characteristic novel feature of this
regime is that the theory no longer contains operators with conformal weight
$1$. It is worthwhile to further investigate this interesting possibility.

{\bf Acknowledgments} I would like to thank A.~Zamolodchikov for illuminating
discussion and E.~Aldrovandi for useful comments. This work was partially
supported by the NSF grant DMS-95-00557.


\begin{thebibliography}{99}

\bibitem{Pol1} A.M.~Polyakov, Phys.~Lett.~{\bf 103B}, 207 (1981).
\bibitem{Pol2} A.M.~Polyakov, {\it Lecture at Steklov Institute},
Leningrad, 1982, unpublished.
\bibitem{TC} L.A.~Takhtajan, {\it Semi-Classical Liouville Theory, Complex
Geometry of Moduli Spaces and Uniformization of Riemann Surfaces}, in:
{\it New Symmetry Principles in Quantum Field Theory}, Eds J.~Fr\"{o}lich et
al,
Plenum Press, New York and London 1992.
\bibitem{T} L.A.~Takhtajan, Mod.~Phys.~Lett.~{\bf A8}, 3529 (1993);
{\bf A9}, 2293 (1995).
\bibitem{TV} L.A.~Takhtajan, {\it Topics in Quantum Geometry of Riemann
Surfaces: Two-Dimensional Quantum Gravity}, to appear in Proceedings of
E.~Fermi Summer School, Varenna 1994; hep-th/9409088.
\bibitem{FS} D.~Friedan and S.~Shenker, Nucl.~Phys.~{\bf B281}, 509 (1987).
\bibitem{ZT} P.G.~Zograf and L.A.~Takhtajan, Funct.~Anal.~Appl.~{\bf 19},
219 (1986); Math.~USSR Sbornik {\bf 60}, 143 (1988); {\bf 60}, 297 (1988).
\bibitem{K} F.~Klein, Math.~Ann.~{\bf 21}, 201 (1883).
\bibitem{P1} H.~Poincar\'{e}, Acta Math.~{\bf 4}, 219 (1884).
\bibitem{P2} H.~Poincar\'{e}, J.~Math.~Pures Appl.~{\bf 5} se.~{\bf 4}, 157
(1898).
\bibitem{CT} E.Braaten, T.~Curtright and C.~Thorn, Phys.~Lett.~{\bf 118B}, 115
(1982); Ann.~Phys.~{\bf 147}, 365 (1983).
\bibitem{KPZ} V.~Knizhnik, A.M.~Polyakov and A.B.~Zamolodchikov, Mod.~Phys.~
Lett. {\bf A3}, 819 (1988).
\bibitem{DDK} F.~David, Mod.~Phys.~Lett.~{\bf A3}, 1651 (1988); J.~Distler and
H.~Kawai, Nucl.~Phys.~{\bf B321}, 509 (1989).
\bibitem{ZZ} A.B.~Zamolodchikov and Al.B.~Zamolodchikov, {\it Structure
Constants and Conformal Bootstrap in Liouville Field Theory}, preprint
RU-95-39; hep-th/9506136.
\bibitem{DO} H.~Dorn and H.-J.~Otto, Nucl.~Phys.~{\bf B429}, 375 (1994).
\bibitem{S} N.~Seiberg, Prog.~Theor.~Phys.~Suppl.~{\bf 102}, 319 (1990).
\bibitem{BPZ} A.A.~Belavin, A.M.~Polyakov and A.B.~Zamolodchikov,
Nucl.~ Phys.~{\bf B241}, 333 (1984).
\bibitem{GM} P.~Ginsparg and G.~Moore, {\it Lectures on 2D Gravity and 2D
String
Theory}, preprint LA-UR-92-3479; hep-th/9304011.
\end{thebibliography}
\end{document}